
%
%
%
%
%
%
%
\input harvmac
\def\frac#1#2{{#1\over#2}}
\def \mult {{\cdot}}
\def \V{{\bf {V}}}
\def \S{{\bf {S} }}
\def \Q{{Q}}
\def \Bbb#1{{\bf #1}}

\def \calC{{\cal{C}}}

\def \zC{{\Bbb C}}
\def \slashh{\widehat}

\Title{\vbox{\baselineskip12pt
\hbox{LAVAL-PHY-94-20} }}
{\vbox{ \centerline{ Quantum Integrals of Motion for the Heisenberg}
\bigskip
\centerline{Spin Chain}
}}

\centerline{Marek P. Grabowski and Pierre Mathieu$^*$}
\bigskip

\centerline{ D\'epartement de physique, Universit\'e Laval,
Qu\'ebec, Canada, G1K 7P4 }

\vskip .3in
An explicit expression for all the  quantum integrals of motion for the
isotropic Heisenberg $s=1/2$ spin chain is presented.
The conserved quantities are expressed in terms of a sum over simple
polynomials in spin variables. This construction is direct and independent
of the transfer matrix formalism. Continuum limits of these integrals
in both ferrromagnetic and antiferromagnetic sectors are briefly discussed.

\vskip 0.15in
\vskip 2.5 in
{\noindent $^*$ supported by NSERC }
\Date{}

\sequentialequations
1. Over the years, the study of quantum spin chains
has provided many important results both in the theory of magnetism and
in mathematical physics. The oldest and still one of the most interesting
models of this type is the isotropic (XXX) Heisenberg $s=1/2$ spin chain,
with the hamiltonian
\eqn\hamilt{ H= g \sum_{i\in \Lambda} { \S_i \S_{i+1} }, }
where $\Lambda$ is the spin lattice and $g$ the coupling. The normalization
of the spin variables $S_j^a$ is chosen to be
\eqn\sutwo {[ S_j^a, S_k^b]= 2 i \delta_{jk} \epsilon^{abc} S_k^c,}
(i.e. $S_k^a$   is a  Pauli sigma matrix,
acting nontrivially only on the $k$-th
factor of the tensor product Hilbert space $\bigotimes_{j\in \Lambda} \zC^2$).
The mathematical structure arising from this innocuous hamiltonian is
astonishingly rich. The key feature accounting for
it is quantum integrability,
i.e. the  existence of a complete set of mutually commuting
integrals of motion.

The isotropic Heisenberg chain and its anisotropic generalizations (XXZ, XYZ)
are one of the simplest quantum integrable systems and as such ideal
laboratories for developing techniques for the study of
more complicated models.
Moreover, these lattice models
have quite interesting continuum limits: the quantum nonlinear
Schr\"odinger (NLS) equation in the ferromagnetic sector ($g<0$) and the
quantum Thirring (or equivalently sine-Gordon) model in the antiferromagnetic
regime ($g>0$). We can thus hope that results obtained in  the lattice
case can be transported to these highly nontrivial quantum field theories.
One of the outstanding problems in these continuous theories
is the explicit construction of the conservation laws. With this in mind, we
have initiated a study of the conserved charges in spin chains, looking first
at efficient ways of calculating their explicit forms, and then
deriving their continuum limits. On the first issue, we end up with
remarkably simple and compact expressions for all the
conservation laws of the XXX model, valid for a finite chain with periodic
boundary conditions and  an infinite chain. These are reported below,
following a brief review of previously known results.
The second issue, which has met with less success, is briefly discussed
at the end of this letter.

2.  The quantum charges for a spin chain are usually defined by means of the
transfer matrix $T$, which  for the XXX model,
is a function of a single spectral parameter $\lambda $.
Building on the Lieb's solution of the two-dimensional classical
ice-type models \ref\Lieb{E. Lieb, Phys. Rev. Lett. {\bf 18}, 692 (1967).},
Sutherland \ref\Suth{B. Sutherland, J. Math. Phys. {\bf 11}, 3183 (1970).}
showed that $[H, T(\lambda)]=0$.  Independently, Baxter \ref\Baxttr{
R. J. Baxter, Ann. Phys. {\bf 70}, 323 (1972).}, in a more general context,
proved the key property of the transfer matrix:
${[T(\lambda), T(\mu)]=0}$.
This implies that the logarithmic derivatives of $T$,
\eqn\defQ {\Q_n\equiv 2 i \frac{ d^{n-1}}{d\lambda^{n-1}}
 \ln T^{-1}(\lambda_0)
T(\lambda)|_{\lambda=\lambda_0}  ,}
where $\lambda_0=i/2 $ in the usual parametrization
\ref\Takht{ L. D. Faddeev and L. A. Takhtajan, J. Sov. Math. {\bf 24}, 241
(1984).}, mutually commute: \eqn\comQ {[Q_n, Q_m]=0.}
The hamiltonian is related to $Q_2$ by $H=g Q_2 + const$; higher charges
correspond to hamiltonians with more neighbors interacting.
As shown by L\"uscher \ref\Lush {M. L\"uscher, Nucl.Phys. {\bf B117}, 475
(1976). }, these charges are local operators, that is, they can be put in the
form:
\eqn\GT{ Q_n=\sum_{\{ i_1, \dots, i_{n-1} \} }
G_{n-1}^T(i_1, \dots, i_{n-1}),}
where the summation is over ordered subsets $\{i_1,\dots,i_{n-1}\}$ of
$\Lambda$, and $G^T$ is a translationally covariant and
totally symmetric function, obeying the locality property:
\eqn\locpro {G_n^T(i_1,\dots,i_n)=0,\quad {\rm for}\quad |i_n-i_1|\ge n.}
Some
additional properties of the XXX charges, in particular  their completeness,
have been proved in
\ref\Babbit{ D. Babbit, L. Thomas, J. Math. Anal. {\bf 72}, 305 (1979).}.
Although the integrals are implicitly known,
it is difficult to extract explicit formulae from
\defQ, even by using computer programs for symbolic computations, since the
size of the transfer matrix grows exponentially with the length of the chain.

There exists however a shortcut. The boost operator,
given by the first moment
of the hamiltonian (and which turns out to be  the derivative of the
logarithm of the Baxter's corner transfer matrix
\ref\BaxCTM{ R. J. Baxter, {\it Exactly Solved Models
in Statistical Mechanics} (Academic Press, London, 1982).}),
\eqn\boost{ B=\frac{1}{2 i} \sum_{j \in \Lambda}j\; \S_j \S_{j+1},}
has been shown \ref\SogoWa{H. B. Thacker, Physica {\bf 18D}, 348 (1986);
H. Itoyama and  H. B. Thacker, Phys. Rev. Lett.
{\bf 58}, 1395 (1987); K. Sogo and  M. Wadati, Prog. Theor. Phys. {\bf 69},
431 (1983).  } to obey
\eqn\BT{ [ B, T(\lambda)]= \frac{\partial}{\partial\lambda} T(\lambda).}
This immediately implies that, up to an additive constant,
\eqn\bHn{ [B, Q_n]= Q_{n+1}.}
$B$ is thus a master-symmetry (see \ref\msymasp{ M. G. Tetelman,
Sov. Phys. JETP {\bf 55}, 306 (1981); B. Fuchssteiner, in
{\it Applications of Field Theory to Statistical  Mechanics,} Lecture
Notes in Physics, vol. 216 (Springer, Berlin, 1985);
E. Barouch and B. Fuchssteiner,
Stud. Appl. Math. {\bf 73}, 221 (1985); E. K. Sklyanin, in
{\it Quantum Groups and Quantum Integrable Systems} (World Scientific,
Singapore, 1992).}).
Eq.  \bHn\ provides a convenient way for a recursive calculation of conserved
charges for an infinite chain.
Using \bHn\ we get, up to additive constants:
\eqn\firstH{ \eqalign { \Q_3=&\sum_{j\in \Lambda} ( \S_j\times \S_{j+1})
\cdot \S_{j+2},\cr
\Q_4=& 2 \sum_{j\in\Lambda}(( \S_j\times \S_{j+1})\times
\S_{j+2})\cdot \S_{j +3} + \S_j\cdot \S_{j+2} - 4 \Q_2,\cr
\Q_5=& 6 \sum_{j\in\Lambda}(((\S_j\times\S_{j+1})\times
\S_{j+2})\times \S_{j+3})\cdot \S_{j+4}
+(\S_j\times \S_{j+2})\cdot \S_{j+3}\cr
&+ (\S_j\times \S_{j+1})\cdot \S_{j+3} - 18 \Q_3.\cr}}
Notice that for a finite chain there appear additional boundary terms in \bHn.
Nevertheless, the expressions \firstH\ are also valid in this case,
if addition in $\Lambda$ is understood modulo $N$, where $N$ is the
number of spins. These expressions
suggest a natural pattern for the structure of the quantum
charges, which we describe and prove below.

3. Before proceeding further, we need to introduce some
notation. A sequence of $n>2$ spin variables,
${\calC=\{\S_{i_1}, ..., \S_{i_n}\}}$,
with $i_1<i_2<...<i_n$, will be called  a cluster of order $n$;
if the ordering  condition is not met,  the
sequence will be called a disordered cluster.
Clusters of a given order can be further classified by specifying their
``holes", that is the sites between  $i_1$ and  $i_n$
that are not included in $\calC $. The number of holes in $\calC$
is clearly  $k=i_n-i_1+1-n$.
Obviously, $k=0$ for a cluster containing only contiguous spins.
We denote as  ${\calC}^{(n,k)}$ the set of all clusters of $\Lambda$ of
order $n$ with $k$ holes. For instance, $\calC^{(3,1)}$ contains
$\{\S_1,\S_2,\S_4\}$, $\{\S_1,\S_3,\S_4\}$ and all their translations.

For any sequence of spins $\calC $, we define $\V_m(\calC)$ as the
vector product of the first $m$ spins, with products nested toward the left,
e.g.
\eqn\jj {\eqalign{&\V_1 =\S_{i_1},\cr
&\V_2=\S_{i_1}\times \S_{i_2},\cr
&\V_3=(\S_{i_1}\times \S_{i_2})\times \S_{i_3},\cr &\dots \cr
&\V_{m+1}=\V_ m \times \S_{i_m}.}}
  {}From these vectors,
we construct scalar $n$-linear polynomials in spin variables
\eqn\fn {f_n(\calC)=\V_{n-1}\mult \S_{i_n}.}
In particular, one has
\eqn\ff{\eqalign{ &f_0=f_1=0,\;f_2=\S_{i_1}\mult \S_{i_2}, \cr
&f_3=(\S_{i_1} \times \S_{i_2})\mult \S_{i_3},\; f_4=((\S_{i_1}\times
\S_{i_2})\times \S_{i_3})\mult \S_{i_4}.}}
The $f_n$'s satisfy an interesting property which is that
the dot product can be placed at  an arbitrary position,
provided that parentheses to its left (right)
are nested toward the left (right), e.g:
\eqn\exnest{\eqalign{f_5 &=((( \S_{i_1}\times \S_{i_2})\times
\S_{i_3})\times \S_{i_4})
\mult \S_{i_5}= (( \S_{i_1}\times \S_{i_2})\times \S_{i_3})\cdot (\S_{i_4}
\times \S_{i_5})\cr &= \S_{i_1} \cdot ( \S_{i_2}\times (\S_{i_3}
\times (\S_{i_4} \times \S_{i_5}))).  }}
This is a direct consequence of the familiar vector identity:
\eqn\famid{ ( {\bf A}\times {\bf B})\cdot {\bf C}={\bf A}\cdot
({\bf B}\times {\bf C}).   }
Finally, we define
\eqn\Fnk{ F_{n,k}=\sum_{\calC \in \calC^{(n,k)}} f_n(\calC).}

The conserved charges can be expressed in a very simple way as
linear combinations of the quantities $F_{n,k}$. It is easily seen that
$ F_{2,0}=\Q_2$ and $F_{3,0} = \Q_3$.
For $n>3$ the charges $Q_n$ obtained from  \bHn\
contain terms proportional to lower order charges.
It will be more convenient to
express the charges in a transformed basis, denoted $\{H_n\}$, in which
these lower order contributions are stripped off.
Our explicit expression for $H_n$ can be most simply
visualized as the sum of the vertices of the tree in \fig\tree{
The tree structure corresponding to $H_n$. The tree stops with the terms
$F_{2,\ell}$ ($F_{3,\ell}$) when $n$ is even (odd). },  with all
vertices contributing with unit weight.
In particular, one has
\eqn\Hnex {\eqalign{ &H_2=F_{2,0}=g^{-1} H, \cr
&H_3=F_{3,0},\cr
&H_4=F_{4,0}+F_{2,1},\cr
&H_5=F_{5,0}+F_{3,1},\cr
&H_6=F_{6,0}+F_{4,1}+F_{2,2}+ F_{2,1},\cr
&H_7=F_{7,0}+F_{5,1}+F_{3,2}+ F_{3,1}, \cr
&H_8=F_{8,0}+F_{6,1}+F_{4,2}+ F_{4,1}+ F_{2,3}+ 2 F_{2,2}+2 F_{2,1},\cr
&H_9=F_{9,0}+F_{7,1}+F_{5,1}+F_{5,2}+ F_{3,3} + 2 F_{3,2}+2 F_{3,1},\cr
&H_{10}=F_{10,0}+F_{8,1}+F_{6,2}+ F_{6,1}+ F_{4,3}+ 2 F_{4,2}+2 F_{4,1}
+F_{2,4}+ 3 F_{2,3}+5 F_{2,2} +5 F_{2,1}.}}
Note that the trees describing $H_{2m}$ and $H_{2m+1}$ have identical
structures.

The algebraic  translation of this construction yields the general expression:
\eqn\Hndef{ H_n= F_{n,0}+ \sum _{k=1}^{[n/2]-1} \sum_{\ell=1}^{k}
\alpha_{k,\ell} F_{n-2k,\ell},}
where the square bracket indicates integer part and the coefficients
$\alpha_{k,\ell}$ are defined  via the recurrence relation:
\eqn\alphadef{ \alpha_{k+1,\ell}=\sum_{m=\ell-1}^k \alpha_{k,m},}
with $\alpha_{1,1}=1$ and $\alpha_{k,0}=0$.
Notice that $\alpha_{k,1}=\alpha_{k,2}$ for $k\ge 2$.
The recurrence relation \alphadef\ can be rewritten in the form:
\eqn\recalpha{ \alpha_{k,\ell}=\alpha_{k-1,\ell-1} +\alpha_{k,\ell+1},}
with the understanding that  $\alpha_{k,\ell}=0$ if $\ell>k$.
This is the defining relation for the generalized Catalan numbers,
${ \alpha_{k,\ell}=C_{2k-l-1, \ell},}$ with $C_{n,m}$ given by
\eqn\Catalrel{ C_{n,m}=\left( {n-1 \over p}\right) -
\left( { n-1 \over p-2}\right),} where
$\left( {a \over b }\right)$ are the binomial
coefficients, with
$p=[(n-m+1)/2]$, $m+n$ odd and $m<n+2$.
In particular, $\alpha_{k,1}=C_{2k,1}$ are the usual Catalan numbers.

For the XXX chain of length $N$ with periodic boundary conditions, the
construction above yields $N-1$ charges $\{ H_2,\dots H_N \} $, which are
clearly independent of each other. To complete this set we may take any of the
three components of the total spin, ${ H_1^a=\sum_{j\in \Lambda} S^a_j }$.
Charges for the infinite XXX  chain are
similarly given by the sequence $\{ H_1^a, H_2,...,   H_n,... \}$.

4. Below we sketch our proof that $\{ H_n \} $ is a family of
conserved charges in involution.
(A complete argument will be published elsewhere.)  First, we note that
since $F_{n,k}$ are invariant under global spin rotation, $[H_1^a, H_n]=0$.
Next we will show that ${[ H_2, H_n]=0}$,
by evaluating directly the commutators:
\eqn\comlink{ [H_2, f_n(\calC)]=\sum_{j\in\Lambda}
[\S_j \S_{j+1}, f_n(\calC)].}
Remarkably, this commutator contains only terms expressible in terms
of the polynomials $f$, namely
$f_{n\pm 1}(\calC^\prime)$, where $\calC^\prime$ can be obtained from
$\calC=\{\S_{i_1},\dots,\S_{i_n}\}$ using a few simple transformations:
\eqn\rulescom{\eqalign{
\S_{i_1-1} \calC &\equiv \{\S_{i_1-1}, \S_{i_1},\dots,\S_{i_n}\},\cr
\calC\S_{i_n+1} &\equiv \{\S_{i_1},\dots,\S_{i_n}, \S_{i_n+1}\},\cr
\calC_{\slashh {i_k}} &\equiv \{ \S_{i_1}, \dots, \S_{i_{k-1}},
\S_{i_{k+1}},\dots,
\S_{i_n}\},\cr
\calC_{ \S_{i_j}\to \S_{i_k} \S_{i_\ell} } &\equiv \{ \S_{i_1},\dots,
\S_{i_{j-1}}, \S_{i_k}, \S_{i_\ell},\S_{i_{j+1}},\dots,\S_{i_n}\},\cr}}
with the last operation being  defined only if $\S_{i_k}, \S_{i_m}$ are
not in $\calC_{ \slashh {i_j}}$.
For $n<N$ the calculation gives:
\eqn\comHfnC{  [ H_2, f_n(\calC)]= a_{n+1,k}(\calC) +
b_{n-1,k+1}(\calC)+d_{n+1,k-1}(\calC)+e_{n-1,k}(\calC)+r(\calC) ,}
where
\eqn\aofC{ a_{n+1,k}(\calC)= - 2 i f_{n+1} (\S_{i_1-1} \calC)+
2 i f_{n+1} (\calC \S_{i_n+1}) ,}
\eqn\bofC{\eqalign{ b_{n-1,k+1}(\calC)= & -4 i
f_{n-1}( \calC_{\slashh {i_2}})
 \delta_{i_1+1,i_2} +4 i f_{n-1}(\calC_{\slashh {i_{n-1}}})\delta _{i_{n-1}+1,
i_{n}} \cr & + 2 i \sum_{j=2}^{n-2} [ f_{n-1}(\calC_{\slashh {i_j}})-
f_{n-1}( \calC_{\slashh{i_{j+1}}}) ] \delta_{i_{j+1},i_j+1} ,}}
\eqn\dofC{ d_{n+1,k-1}(\calC)=2 i f_{n+1}( \S_{i_1} \S_{i_1+1}
\calC_{\slashh {i_1}}) (1-\delta_{i_1,i_2-1}) - 2 i
 f_{n+1}(\calC_{\slashh {i_n}}\S_{i_n-1} \S_{i_n})
(1-\delta_{i_n, i_{n-1}+1}), }
 \eqn\eofC{ e_{n-1,k}(\calC)=4 i f_{n-1}
( \calC_{\slashh {i_1}}) \delta_{i_1+1,i_2}-
4 i f_{n-1}( \calC_{\slashh {i_n}})\delta_{i_{n-1}+1,i_n}, }
 \eqn\rofC{ r(\calC)=-2 i \sum_{j=2}^{n-2} f_{n-1}( \calC_{\S_{i_j}\to
\S_{i_j}\S_{i_j+1}}) (1-\delta_{i_{j+1}, i_j+1})
+2 i \sum_{j=3}^{n-1} f_{n+1}(\calC_{\S_{i_j}\to \S_{i-1}\S_i})
 (1-\delta_{i_{j-1}, i_j-1}).  }
The reader is cautioned to distinguish between
$\S_{i_j+1}$ and $\S_{i_{j+1}}$
in the above formulae. Note that, with the exception of $r(\calC)$, all of
the terms on the left hand-side of \comHfnC\ involve only ordered clusters.
 Summing up over all possible clusters, we get:
\eqn\coHFnk{ [ H_2, F_{n,k}]=A_{n+1,k}+D_{n+1,k-1}+ B_{n-1,k+1} ,}
where $A_{n+1,k}=
\sum_\calC a_{n+1,k}(\calC)$, $B_{n-1,k+1}=\sum_\calC b_{n-1,k+1}(\calC)$,
$D_{n+1,k-1}=\sum_\calC d_{n+1,k+1}(\calC) $.
Due to symmetry $\sum_\calC e_{n-1,k}(\calC)=\sum_{\calC} r(\calC)=0$
(hence contributions from disordered clusters cancel).
In the case $n=N$ the calculation yields:
\eqn\HFnzer{ [ H_2,F_{N,0}]=B_{N-1,1}.}
It follows from \coHFnk\ and \HFnzer, that in order to prove that
$H_n$ commute with  the hamiltonian, it is sufficient to show that
for any $n\le N$,

(i) $[ H_2, F_{n,0}]$ does not contain terms of order $n+1$,

(ii) $[ H_2, F_{n,k}+\sum_{\ell=1}^{k+1} F_{n-2,\ell}]$ does not
contain terms of order $n-1$.

{\noindent  {
The assertion (i) is immediate. If $n=N$, it follows from \HFnzer.
If $n < N$, it follows from the fact that, due to translational symmetry,
${ A_{n+1,0}=0}$.  (ii) is equivalent to:
\eqn\comtwo{ B_{n-1,k+1}+\sum_{\ell=1}^{k+1}
 (A_{n-1,\ell}+ D_{n-1,\ell-1} )=0.}
The above sum contains contributions of clusters of order $n-1$, with
hole numbers ranging from 0 to $k+1$; thus the sum vanishes if and
only if the individual contributions vanish, which can be proved by  a
tedious calculation using \aofC-\dofC.
}\indent}

Having established that the $H_n$'s commute with the
hamiltonian, we have yet to show that they commute among themselves.
We also want  to express the logarithmic derivatives of the transfer
matrix in the basis $\{H_n\}$.
To this end, we calculate the commutators of $H_n$ ($n\ge2$) with the boost
operator:
\eqn\coBHn{ [ B, H_n]=\sum_{m=0}^{\max(1,[n/2]-1)} \beta^{(n)}_m H_{n+1-2m}, }
where
\eqn\betacoef{ \beta^{(n)}_0=n-1,
\;\;\beta^{(n>2)}_1=5-3n,\;\;
\beta^{(n)}_{1<\ell < [n/2]}=-(n-2 \ell-1) \alpha_{\ell,1} .}
{} From \coBHn\ it is clear that $Q_n$ of even (odd) order $n$ can be
expressed as a linear combination of the $H_m$ with even (odd) $m\le n$:
\eqn\QnHn{ \Q_n=\sum_{p=0}^{[n/2]-1}\gamma^{(n)}_p H_{n-2p}.}
The coefficients $\gamma$ satisfy the recurrence relation:
\eqn\gammarec{\gamma_{\ell}^{(n+1)}=\sum_{p,m\ge 0 \atop p+m=\ell}
\gamma_p^{(n)}\beta_{m}^ {(n-2p)}, } with $\gamma^{(2)}_p=\delta_{p,0}$.
In particular, modulo additive constants,
\eqn\QninHn{\eqalign{ & Q_4=2 H_4 - 4 H_2,\cr
&Q_5=6 H_5-18 H_3,\cr
&Q_6=24 H_6- 96 H_4 + 72 H_2.\cr}}
Since $\{Q_n\}$ form a family of conserved charges in involution,
(cf. \comQ),
it then follows from \QnHn\ that all of the $H_n$ mutually commute.
This can be also   proved directly, without using \comQ,
by an inductive argument based on  \coBHn\ and the Jacobi identity.

5. The ferromagnetic sector of the XXX chain with an arbitrary spin $s$
has a nonrelativistic dispersion relation, which can be brought to light by
means of the  Holstein-Primakoff transformation
\ref\HolstPr{ T. Holstein and H. Primakoff, Phys. Rev.
{\bf 58}, 1098 (1940).}.
Under this transformation the chain is mapped to a lattice version of the
quantum nonlinear Schr\"odinger (NLS) equation \ref\Tara{A. G. Izergin and
V. E. Korepin, Nucl. Phys. {\bf B205}, 401 (1982);
V. O. Tarasov, L. A. Takhtajan, and L. D. Faddeev, Theor. Math. Phys.
{\bf  57}, 1059 (1984).}, whose continuous hamiltonian is
\eqn\HNLS{ H_{NLS}=\int_{-\infty}^{\infty} (\Psi^+_x \Psi_x +
\kappa \Psi^+\Psi^+\Psi\Psi) dx,}
with $[\Psi^+(x), \Psi(y)]=\delta (x-y)$. The coupling constant $\kappa$ is
related to the spin $s$ by $\kappa=-2/(s \Delta)$, $\Delta$ being the lattice
constant. Finding explicit forms of the conserved charges in
this system is an interesting open
problem, which has been recently investigated by several authors \ref\Schrch{
K. M. Case, J. Math. Phys. {\bf 25}, 2306 (1984); E. Gutkin,
Ann. Inst. Henri Poincar\'e {\bf 2}, 67 (1985); B. Davies, Physica
{\bf 167A}, 433 (1990).}.
To make contact with our results for the $s=1/2$ chain, one must take the
limit $\kappa\to \infty$ (the impenetrable boson system).
In this limit the conserved charges have the form:
\eqn\Impbos{\eqalign{ H_{2n}&\to
\int_{-\infty}^\infty \sum_{m=1}^{n} a^{(n)}_m\;\;
(\Psi^+)^m (x)\Psi^m(y) \delta^{n-m+1} (x-y) dx dy,\cr
H_{2n+1}&\to \int_{-\infty}^\infty \sum_{m=1}^{n} b^{(n)}_m\;\;
(\Psi^+)^m(x)\Psi_y(y)
\Psi^{m-1}(y) \delta^{n-m+1}(x-y) dx dy ,\cr}}
where $\delta^k$ denotes  a suitably regularized $k$-th
power of the Dirac delta
function. (An extended discussion of these results will be published
elsewhere.) The continuum limit of the $s=1/2$ XXX charges using the
Holstein-Primakoff transformation, does indeed lead to
integrals of this type. However, in general the limiting process is
ambiguous as far as ordering of operators is concerned (e.g. on the lattice
$\Psi_i\Psi_j^+= \Psi_j^+\Psi_i$ if $i\ne j$, but these two expressions have
different continuous limits). We have not yet found a prescription
which fixes
these ambiguities and reproduces the values of the coefficients in \Impbos.

In the antiferromagnetic sector, the continuum limit of the $s=1/2$ XXX spin
chain is a special case of the massive Thirring model
\ref\Luther{A. Luther and B. Peschel, Phys. Rev. B {\bf 12}, 3908 (1975). },
related to the sine-Gordon model through bosonization
\ref\Coleman{S. Coleman, Phys. Rev. D {\bf 11}, 2088 (1975).}. It can be
equivalently described by a level one $SU(2)$  Wess-Zumino-Novikov-Witten
(WZNW) model with a marginal perturbation
\ref\Affleck{ I. Affleck, Phys. Rev. Lett. {\bf 55}, 1355 (1985);
Nucl. Phys. {\bf B265}, 409 (1986).}.
In the framework of the latter model,
by  replacing $S^a_j$ by the the current
algebra generators
$J^a(x_j)$, and taking the lattice constant $\Delta\to 0$, one gets
the holomorphic part of the conserved charges:
$H_2\to \int T dx $, $H_3\to 0$ (this generalizes to all $H_{2m+1}$),
$H_4\to \int (T T) dx$, where $T$ is the Sugawara
energy-momentum tensor and the
parenthesis denotes the standard  normal ordering in conformal field theory.
These conserved integrals are common to the quantum KdV equation
\ref\Pierre{B. Kuperschmidt and P. Mathieu,
Phys. Lett. {\bf B227}, 245 (1989);
R. Sasaki and I. Yamanaka, Adv. Stud. Math.  {\bf 16}, 271 (1988);
T. Eguchi and S. Kang, Phys. {\bf B224}, 373 (1989). }
and, via a Feigen-Fuchs transformation, the quantum sine-Gordon equation.
However, calculation of the limits of the higher order charges,
needed to probe the value of the WZNW central charge $c=1$,
is again plagued by
ordering ambiguities.

6. In this work we have presented a simple and compact expression for the
conservation laws of the $s=1/2$ XXX Heisenberg chain.
Our construction of the conserved charges is
independent of the transfer matrix formalism and uses only
the algebraic relations \coHFnk, \HFnzer, and \comtwo. It thus provides
an alternative and direct way of proving the integrability of the XXX chain.
It is an interesting question whether this construction can be generalized to
the anisotropic case or to other integrable spin chains.

After the completion of this work, we became aware of
\ref\Ansh { V. V. Anshelevich, Theor. Math. Phys. {\bf 43}, 350 (1980).},
which presents an expression for the basis of the space of quantum integrals
of motion of the infinite XXX chain,   given without proof nor with any
indication of how it has been found. This basis is different from ours but
we have checked that the two yield equivalent results. However, the
 logarithmic derivatives of the transfer matrix have not been
given in \Ansh.

\listrefs \listfigs \bye